\documentclass[final,3p,times,twocolumn]{elsarticle}

\usepackage{here}
\usepackage{graphicx,graphics}
\usepackage{epsfig}
\usepackage{amssymb,amsthm,amsmath,amsfonts}
\usepackage{bbold}

\journal{Nuc. Phys. (Proc. Suppl.)}

\begin{document}

\begin{frontmatter}

\title{Anomalous $AV^*V$ vertex in the soft-wall holographic model of QCD}
\author[label1]{Stefano Nicotri\corref{cor1}}
\address[label1]{Universit\`a degli Studi di Bari \& INFN, Sezione di Bari\\
via Orabona~4, I-70125, Bari, Italy}
\cortext[cor1]{Speaker}
\ead{stefano.nicotri@ba.infn.it}

\begin{abstract}
\noindent
We discuss the vertex function of two vector and one axial-vector operators in the soft-wall holographic model of QCD. 
When one of the two vector currents represents an on-shell soft photon, such a vertex is described by two structure functions $w_L$ and $w_T$, which are usually calculated through triangular loop diagrams. 
We evaluate these functions in the soft-wall model of holographic QCD (HQCD) and compare the outcome to the QCD findings. 
\end{abstract}


\end{frontmatter}

The analysis of processes involving nuclear and subnuclear interactions has been puzzling physicists since long time, being perturbation theory not reliable for this class of phenomena.
This is due to the nonperturbative nature of QCD, the theory describing hadrons, at low energy.
Among the many approaches that have been developed to face this problem, the one related to the AdS/CFT (Anti de~Sitter/Conformal Field Theory) correspondence (or gauge gravity duality) \cite{maldacena1,Witten:1998qj,gubser-klebanov-polyakov} is currently being looked at with increasing interest among theoretical physicists.
The key concept underlying this approach is that the strong coupling regime of QCD (in the limit of large number of colours $N_c$) is conjectured to be dual to the semiclassical limit of some higher dimensional theory (or model) including gravity (thus defined in a curved spacetime).
This is attractive, since complicated nonperturbative calculations involving the strong interaction could be carried out in the simpler dual (perturbative) gravity theory and the results ``translated'' back in the real world.
With this concept in mind, two main directions have been followed with the aim of finding the gravity dual of QCD and constructing such a dictionary. 
Both have to face the problem that QCD is neither supersymmetric nor conformal (as the CFT involved in AdS/CFT, $\cal N$=4~SYM).
Here we choose to follow a bottom-up approach, by using a model constructed \emph{ad hoc} to reproduce some desirable features of QCD: the soft-wall holographic model of QCD \cite{andreev1,Karch:2006pv}.

Here we want to investigate the $AV^*V$ vertex in this holographic framework.
Let us consider the three-point correlation function involving two vector currents $J_\mu=\bar qV\gamma_\mu q$ and an axial-vector current  $J^5_\nu=\bar qA\gamma_\nu\gamma_5 q$, with quark fields $q^i_f$ carrying a colour $(i)$ and a flavour $(f)$ index, and $V$ and $A$ diagonal matrices acting on the flavour indices:
\begin{eqnarray}\label{threepoint}
 T_{\mu\nu\sigma}(q,k) & = & i^2\,\int d^4x\,d^4y\,e^{i\,q\cdot x-i\,k\cdot y}\times\\
 & \times & \langle0| T[J_\mu(x)J_\nu^5(0)J^{em}_\sigma(y)] |0\rangle\,\,.\nonumber
\end{eqnarray}
Taking the limit in which one vector current represents a real and soft photon (i.e. with four-momentum $k\simeq0$, $k^2=0$, and polarization $\epsilon^\alpha$), \eqref{threepoint} is related to the two-point function of $J_\mu$ and $J^5_\nu$ in an external electromagnetic field
\begin{equation}\label{twopoint}
T_{\mu \nu}(q,k)=i\,\int d^4x\,e^{i\,q\cdot x}\langle0|T[J_\mu(x)J_\nu^5(0)]|\gamma(k,\epsilon)\rangle
\end{equation}
by the relation $T_{\mu \nu}(q,k)=e\,\epsilon^{\sigma}\,T_{\mu\nu\sigma}(q,k)$.
In this limit, $T_{\mu \nu}$ can be expressed in terms of two structure functions $w_L(q^2)$ and $w_T(q^2)$, longitudinal and transversal with respect to the axial current index, respectively:
\begin{eqnarray}\label{decomp}
T_{\mu\nu}(q,k) & = & -\frac{i}{4\pi^2}{\rm Tr}\left[QVA\right]\left\{w_T(q^2)(-q^2 {\tilde f}_{\mu\nu}+\right.\nonumber\\
&& \left.+q_\mu q^\lambda{\tilde f}_{\lambda \nu}-q_\nu q^\lambda{\tilde f}_{\lambda\mu})+\right.\\
&& \left.+w_L(q^2)q_\nu q^\lambda{\tilde f}_{\lambda \mu}\right\}\,,\nonumber
\end{eqnarray}
where $Q$ is the electric charge matrix and ${\tilde f}_{\mu \nu}=\epsilon_{\mu\nu\alpha\beta} f^{\alpha\beta}/2$ is the dual field of the photon field strength $f^{\alpha\beta}=k^\alpha\epsilon^\beta-k^\beta\epsilon^\alpha$.
The correlators \eqref{threepoint} and \eqref{twopoint} involve anomalous triangle diagrams, thus the structure functions $w_L$ and $w_T$ carry information about the QCD anomaly.
When such diagrams involve just one fundamental quark of mass $m$, the one-loop result for $T_{\mu\nu}$ at large Euclidean $Q^2$ is \cite{Adler:1969gk}
\begin{equation} \label{wLT-1loop}
w_L(Q^2)=2w_T(Q^2) =\frac{2N_c}{Q^2}\left[1+2\frac{m^2}{Q^2}\ln\frac{m^2}{Q^2}+{\cal O}\left(\frac{m^4}{Q^4}\right)\right]\,.
\end{equation}
A non-renormalization theorem for the anomaly states that $w_L$ does not receive perturbative corrections \cite{Adler:1969er}, while the perturbative corrections to $w_T$ vanish to all orders in the limit $k\to0$ and $Q^2 \gg m^2$ \cite{Vainshtein:2002nv}. 
This means that in the chiral limit $m=0$ there are no $\alpha_s$ corrections to both  $w_L$ and $ w_T$, and the functions are completely determined by the anomalous triangle diagrams: $w_L(Q^2)=2N_c/Q^2$ and (not considering nonperturbative corrections) $w_L(Q^2)=2w_T(Q^2)$.
$w_L$ does not receive nonperturbative corrections, while $w_T$ gets the first nonperturbative contribution at ${\cal O}(1/Q^6)$ (in the chiral limit):
\begin{equation}\label{wTchiralQCD}
w_T(Q^2)=\frac{N_c}{Q^2}+\frac{128\pi^3\alpha_s\chi\langle\bar qq\rangle^2}{9Q^6}+{\cal O}\left(\frac{1}{Q^8}\right)
\end{equation}
where $\langle\bar qq\rangle$ is the vacuum quark condensate and $\chi$ its magnetic susceptibility, introduced to parametrize the tensor condensate
\begin{equation}\label{magnsusc}
\langle0|\bar q\sigma^{\rho\tau}q|\gamma(k,\epsilon)\rangle=ie\langle\bar qq\rangle\chi\,f^{\rho \tau}\,.
\end{equation}
Keeping $m$ finite instead (and using \eqref{magnsusc}), we have
\begin{eqnarray}
w_L(Q^2) & = & 2w_T(Q^2)=\frac{2N_c}{Q^2}\left[1+\frac{2m^2}{Q^2}\ln\frac{m^2}{Q^2}-\right.\nonumber\\
& - & \left.\frac{8\pi^2m\langle\bar qq\rangle\chi}{N_c Q^2}+{\cal O}\left(\frac{m^4}{Q^4} \right)\right]
\end{eqnarray}
the relation between $w_L$ and $w_T$ standing up to ${\cal O}(1/Q^4)$.

We now want to investigate the structure functions in \eqref{decomp} by means of the soft-wall holographic model of QCD \cite{Colangelo:2011xk}.
The model is defined in an AdS$_5$ space with line element
\begin{equation}\label{metric}
ds^2=g_{MN}dx^M dx^N=\frac{R^2}{z^2}(\eta_{\mu \nu}dx^\mu dx^\nu-dz^2)\,.
\end{equation}
The coordinate indices $M,N$ are $M,N=0,1,2,3,5$,  $\eta_{\mu\nu}=\mbox{diag}(+1,-1,-1,-1)$ and $R$  is the AdS$_5$ curvature radius (set to unity from now on).
In the model, the fifth ``holographic'' coordinate $z$ runs in the range $\epsilon\leqslant z<+\infty$, with $\epsilon\to0^+$ and it takes the meaning of an inverse energy scale, with  $\epsilon$ playing the role of a renormalization scale.
The five dimensional action defining the model is
\begin{eqnarray}\label{action}
S & = & S_{YM}+S_{CS}\nonumber\\
S_{YM} & = & \frac{1}{k_{YM}}\int d^5x\,\sqrt{g}\,e^{-\Phi}{\rm Tr}\biggl\{|DX|^2-m_5^2 |X|^2-\nonumber\\
& - & \frac{1}{2 g_5^2} (F_V^2+F_A^2)\biggr\}\,.\\
S_{CS} & = & 3k_{CS}\,\epsilon_{ABCDE}\int d^5x{\rm Tr}\left[{A}^A \left\{F_{V}^{BC}, F_{V}^{DE}\right\}\right]\label{actionCS}\nonumber
\end{eqnarray}
$S_{YM}$ is the Yang-Mills action involving a vector and an axial vector $U(N_f)_V\otimes U(N_f)_A$ gauge fields $V_M=V_M^aT^a$ and $A_M=A_M^aT^a$, dual to the dimension-three vector and axial currents, respectively, with $T^a$ the generators of the $U(N_f)$ Lie algebra ($a=0,\ldots,N_f^2-1$\footnote{$T^0=\frac{\mathbb{1}}{\sqrt{2N_f}}$ and ${\rm Tr}(T^aT^b)=\delta^{ab}/2$}).
The corresponding strength tensors are
\begin{eqnarray}
F_V^{MN} & = & \partial^M V^N-\partial^N V^M-i[V^M,V^N]-i[A^M,A^N]\nonumber\\
F_A^{MN} & = & \partial^M A^N-\partial^N A^M-i[V^M,A^N]-i[A^M,V^N]\,.\nonumber
\end{eqnarray}
A background dilaton-like field $\Phi(z)=(cz)^2$ is introduced in $S_{YM}$ to break conformal symmetry, mimicking some features of the QCD nonperturbative vacuum.
Its functional form is chosen to obtain linear Regge trajectories for light vector mesons; $c$ is a dimensionful parameter setting a scale for QCD quantities, and its numerical value $c=M_\rho/2$ is obtained from the spectrum of $\rho$ mesons.
$X(x,z)=X_0(z)e^{i\pi(x,z)}$ is a scalar tachyon dual to the quark bifundamental operator ${\bar q}_R^\alpha q_L^\beta$.
$X_0(z)=v(z)/2$ only depends on the holographic coordinate $z$ and is dual to the v.e.v. of the scalar operator, containing information concerning chiral symmetry breaking.
In particular, it is considered as a background field, whose functional form is chosen on a dimensional basis, taking the lowest orders in the $z\to0$ expansion of the solution of the equation of motion and treating the coefficients as external parameters \cite{Kwee:2007dd}
\begin{equation}\label{v}
v(z)=mz + \sigma z^3\,.
\end{equation}
$m$ is dual to the quark mass, governing explicit chiral symmetry breaking, while $\sigma\propto\langle\bar qq\rangle$ \cite{Colangelo:2011sr,Colangelo:2011xk} determines the spontaneous breaking.
The action $S_{YM}$ describes vector, axial, and pseudoscalar mesons, and it has been extended to include also fields describing scalar mesons and glueballs \cite{Colangelo:2007if,Colangelo:2007pt,Colangelo:2008us} which we omit here.
Matching the two-point correlation functions of the vector and of the scalar field with the corresponding leading order perturbative QCD results, the parameters $g_5^2$ and $k_{YM}$ are fixed to the values $g_5^2=3/4$ and $k_{YM}=16\pi^2 /N_c$.
The second part of \eqref{action}, $S_{CS}$, is a Chern-Simons term, a topological term needed to holographically describe the anomaly \cite{Witten:1998qj,Hill:2006wu,Grigoryan:2008up,Gorsky:2009ma,Brodsky:2011xx,Son:2010vc}\footnote{A boundary term has been added to make the invariance under a vector gauge transformation explicit, and higher odd powers of the fields have been omitted since they do not contribute to the $AV^*V$ correlation function.}.
Performing a Fourier transform with respect to the 4d coordinates $x^\mu$ of the gauge fields $V_\mu^a(x,z)$ and $A_\mu^a(x,z)$ (in the axial gauge $V_z=A_z=0$), we can write each field as a product ${\tilde G}^a_\mu(q,z)=G(q,z)G_{\mu 0}^a(q)$, where $G(q,z)$ is the so-called bulk-to-boundary propagator and $G_{\mu0}^a(q)$ is the source of the corresponding dual operator in the QCD generating functional ($G=V,A$).
Moreover, using the parallel and transverse projection tensors $P_{\mu\nu}^\parallel=q_\mu q_\nu/q^2$ and $P_{\mu \nu}^\perp=\eta_{\mu \nu}-P_{\mu\nu}^\parallel$, we can express the bulk-to boundary propagators in terms of the transverse and longitudinal parts\footnote{Notice that the vector field is transverse.}:
\begin{eqnarray}\label{perp-par}
\tilde V^a_\mu(q,z) & = & V_\perp(q,z)P_{\mu\nu}^\perp V_0^{a\nu}(q)\\
\tilde A^a_\mu(q,z) & = & A_\perp(q,z)P_{\mu \nu}^\perp A_0^{a\nu}(q)+A_\parallel(q,z)P_{\mu \nu}^\parallel A_{ 0}^{a\nu}(q)\,,\nonumber
\end{eqnarray}
with boundary conditions $V_\perp(q,0)=A_\perp(q,0)=A_\parallel(q,0)=1$.
To obtain the correlation function we are interested in, we use the AdS/QCD relation:
\begin{equation}\label{adsqcd}
\biggl\langle e^{i\int d^4x\;{\cal O}(x )\,f_0(x)}\biggr\rangle_{QCD}=e^{iS[f(x,z)]}\,,
\end{equation}
identifying the QCD generating functional (on the lhs) with the (semiclassical limit of the\footnote{This consists in taking just the exponential of the five dimensional effective action, not considering the contributions coming from the quantum fluctuations.}) partition function of the higher dimensional dual model (on the rhs).
Eq.~\eqref{adsqcd} represents the general relation involving a QCD operator ${\cal O}(x)$ (the axial and vector currents and the scalar operators in our case) and $S$ is the effective 5d action \eqref{action}. 
The source of ${\cal O}(x)$ coincides with the $z=0$ boundary value $f_0(x)=f(x,0)$ of the dual field $f(x,z)$ in the 5d action.
The correlation function of a vector and an axial vector current in the external electromagnetic background field can be written in terms of the functions $w_L$ and $w_T$:
\begin{eqnarray}\label{JJ-F}
d^{ab}\langle J_\mu^{V} J_\nu^{A} \rangle_{\tilde F} & = & i \int d^4 x\,e^{iqx}\langle T\{J_\mu^{Va}(x)J_\nu^{Ab}(0)\}\rangle_{\tilde F}\nonumber\\
& = & d^{ab}\frac{Q^2}{4\pi^2}P_{\mu\alpha}^\perp\left[P_{\nu\beta}^\perp w_T(Q^2)+\right.\nonumber\\
& + & \left.P_{\nu\beta}^\parallel w_L(Q^2)\right]\tilde F^{\alpha\beta}\,.
\end{eqnarray}
The two terms in this expression, the one proportional to $P_{\mu\alpha}^\perp P_{\nu\beta}^\perp$ and the other one proportional to $P_{\mu\alpha}^\perp P_{\nu\beta}^\parallel$, can be obtained by functional derivation of the on-shell action \eqref{action} \cite{Colangelo:2011xk} and from the comparison of the result with \eqref{JJ-F} ($y=cz$):
\begin{eqnarray} \label{wL-wT}
w_L(Q^2)&=&-{2 N_c \over Q^2}\int_0^{\infty} dyA_\parallel(Q^2,y)  \partial_y V(Q^2,y)\nonumber\,,\\
\\
w_T(Q^2)&=&-{2 N_c \over Q^2}\int_0^{\infty} dy A_\perp(Q^2,y) \partial_y V(Q^2,y)\nonumber\,.
\end{eqnarray}
The coefficient $2 N_c$ has been obtained fixing the factor $k_{CS}$ in $S_{CS}$ to the value $k_{CS}=-\frac{N_c}{96\pi^2}$ from the QCD OPE.
Substituting in \eqref{wL-wT} the solutions of the equations of motion coming from \eqref{action}, we can evaluate the two structure functions by a simple one dimensional integral, for \emph{any} value of $Q^2$.
We now examine the results in different cases.
In the simplest case $m=\sigma=0$ the anomaly equation $w_L=2w_T=2N_c/Q^2$ is recovered.
In the chiral limit $m=0$, $\sigma\neq0$, the large-$Q^2$ behaviour of $w_L$ and $w_T$ can be found analytically:
\begin{eqnarray}
w_L(Q^2) & = & \frac{2N_c}{Q^2}\,,\label{wL-wTchiral1}\\
w_T(Q^2) & = & \frac{N_c}{Q^2}-\tau\,g_5^2\,\sigma^2\,\frac{2N_c}{Q^8}+{\cal O}\left(\frac{1}{Q^{10}}\right)\label{wL-wTchiral2}
\end{eqnarray}
with $\tau=2.74$.
We notice that this result is different from the QCD one \eqref{wTchiralQCD}, where the first correction, proportional to the magnetic susceptivity $\chi$, is of ${\cal O}(1/Q^6)$.
This means that the soft-wall model is incomplete, since the tensor field dual to the $\bar q\sigma^{\rho\tau}q$ operator is not included, and this can be responsible for the difference between \eqref{wL-wTchiral2} and \eqref{wTchiralQCD}.
The inclusion of such an operator in HQCD is still under investigation \cite{Domokos:2011dn}.
If one considers a finite value of the quark mass $m\neq0$, the result
\begin{eqnarray}\label{wL-wTcomplete}
w_T(Q^2) & = & \frac{N_c}{Q^2}\left(1-\frac{g_5^2 m^2}{3 Q^2}-\frac{2g_5^2 m^2 c^2}{5 Q^4}+\frac{g_5^4 m^4}{6 Q^4}-\right.\nonumber\\
& - & \left.\frac{8g_5^2 m\sigma}{5 Q^4} \right)+{\cal O}\left(\frac{1}{Q^8}\right)\,,\nonumber\\
&&\\
w_L(Q^2) & = & \frac{2N_c}{Q^2}-\left[1-\pi(Q^2,0)\right]\,N_c\left[\frac{g_5^2\, m^2}{Q^4} +\right.\nonumber\\
& + & \left.\frac{4g_5^2\,m\,\sigma}{Q^6} - \frac{2g_5^4\,m^4}{3Q^6}+{\cal O}\left(\frac{1}{Q^8}\right)\right]\,\nonumber
\end{eqnarray}
is found, where $\pi(Q^2,0)$ is a boundary condition for the chiral field, an external parameter that needs to be fixed from some external input.
Regardless of this, also here, we notice that the form of the corrections in \eqref{wLT-1loop} is not recovered.
This is probably due to the fact that the assumption \eqref{v} we have made for the v.e.v. of the scalar field is too naive.
A better model, with the v.e.v. arising dynamically from the equations of motion, or at least containing a logarithmic term seems to be required to fix this problem.
Moreover, the relation $w_L=2w_T$ is violated al large $Q^2$.

\section*{Conclusions \& outlook}
We have shown the application of a promising and quite innovative method to the nonperturbative regime of QCD, evaluating the $AV^*V$ vertex structure functions through the holographic soft-wall model.
Many results can be obtained analytically with less efforts than other more involved methods, like lattice QCD, and they can be efficiently compared with experiments and with the outcome of other effective approaches, like QCD sum rules \cite{Shifman:1978bx,Narison:1989aq}, or of lattice field theory.
The method also shows problems in correctly determine the functional form of the corrections to the anomaly relations, problems related to the naivety of the assumptions that have been made.

\section*{Acknowledgements}
\noindent
I thank P.~Colangelo, F.~De~Fazio, F.~Giannuzzi, and J.J.~Sanz-Cillero for collaboration.
I would also like to thank the Organizers of the ``QCD~12'' Conference for the fresh and stimulating environment they have contributed to create.

\end{document}